\documentclass[12pt]{emulateapj}
\pdfoutput=1
\usepackage{blindtext}
\usepackage{float}
\usepackage{placeins}
\usepackage{graphicx}
\usepackage{amssymb, amsmath}
\usepackage{natbib}

\begin{document}
\title{Particle Acceleration and the Origin of X-ray Flares in GRMHD simulations of SGR~A*}
\author{David Ball}
\author{Feryal \"Ozel}
\author{Dimitrios Psaltis}
\author{Chi-kwan Chan}
\affil{Steward Observatory and Department of Astronomy, University of Arizona}

\begin{abstract}
Significant X-ray variability and flaring has been observed from Sgr~A* but is poorly understood from a theoretical standpoint.  We perform GRMHD simulations that take into account a population of non-thermal electrons with energy distributions and injection rates that are motivated by PIC simulations of magnetic reconnection.  We explore the effects of including these non-thermal electrons on the predicted broadband variability of Sgr~A* and find that X-ray variability is a generic result of localizing non-thermal electrons to highly magnetized regions, where particles are likely to be accelerated via magnetic reconnection.  The proximity of these high-field regions to the event horizon forms a natural connection between IR and X-ray variability and accounts for the rapid timescales associated with the X-ray flares.  The qualitative nature of this variability is consistent with observations, producing X-ray flares that are always coincident with IR flares, but not vice versa, i.e., there are a number of IR flares without X-ray counterparts.
\end{abstract}
\maketitle

\section{Introduction}
Observational evidence and theoretical arguments strongly suggest that accretion flows around low-luminosity supermassive black holes are radiatively inefficient (see a recent review by \citealt{yuan2014}).  Current models of radiatively inefficient accretion flows employ  simplified thermodynamic prescriptions of the rarefied plasma in the disk.  This is due to a number of complicating factors that make it difficult to fully describe the physics of the system. 

Low-luminosity accretion flows are composed of two temperature plasmas, where ions are significantly hotter than electrons.  This occurs for several reasons.  First, the very low-density plasma is nearly collisionless, such that thermalization via Coulomb collisions is inefficient.  Second, the electrons are able to radiate away their energy much faster than the ions.  Third, not only do cooling rates differ between the electrons and ions, but the mechanisms responsible for heating the fluid often favor one component of the plasma over the other, with the magnitude of this heating asymmetry between ions and electrons depending on the plasma properties \citep{quataert1999}.  Processes such as the dissipation of turbulent energy, magnetic reconnection, and shocks all play a role in mediating the temperature of the fluid, but the details of these mechanisms and their combined impact on the thermodynamics of low-luminosity accretion flows is not fully understood.  

Simulations of low-luminosity accretion flows generally assume a Maxwellian distribution of electrons at a prescribed temperature (e.g., \citealt{dexter2012,drappeau2013,moscibrodzka2014,chan2015a}).  More recently, \citet{ressler2015} developed a method for independently evolving the electron temperature distribution, taking into account spatially varying heating as well as anisotropic electron conduction, but still assuming a thermal particle distribution.  However, subgrid modeling of heating and acceleration processes indicate that some fraction of particles are likely to be accelerated into a non-thermal distribution.

Astrophysical particle acceleration is a heavily studied field with broad implications for many astrophysical systems (for a recent review, see \citealt{lazarian2012} and references therein).  There have recently been significant improvements in our understanding of the numerous heating and acceleration mechanisms that are relevant to accretion flows from a microphysical standpoint.  By utilizing particle-in-cell (PIC) simulations, \citet{sironi2014} showed that relativistic reconnection generally accelerates the particles in a plasma into power-law like distributions.  \citet{guo2014} used PIC simulations to determine the effect of low Mach number shocks on acceleration and showed that it also produces a non-thermal distribution.  These modeling efforts at small scales have yielded new insight into the fundamental properties of heating and acceleration mechanisms, but their effects have not yet been incorporated, as sub-grid models, into the larger scale general relativistic magnetohydrodynamic (GRMHD) simulations of low-luminosity accretion flows.  By coupling the results of the microphysical models to GRMHD simulations, it will become possible to more robustly interpret some observed phenomena that have thus far eluded a physical explanation.  

The role of non-thermal electron energy distribution has been addressed in the context of stationary hydrodynamic models by \citet{mahadevan1998}, \citet{ozel2000} and, \citet{yuan2003}, who showed that even a relatively small number of power-law electrons can significantly impact the spectra predicted from a model, generating X-ray power-law tails as well as boosting the low frequency radio flux.  These studies used analytical steady-state solutions to calculate spectra, which do not capture short timescale GRMHD effects that play a large role in determining the variability properties of the system.  \citet{dodds-eden2010} used two dimensional time-dependent MHD models and injected non-thermal electrons into regions of rapidly changing magnetic fields in an effort to explain the rapid flares seen from Sgr~A*.  These models, however, do not account for the three dimensional character of the flow, as well as for general relativistic effects such as strong lensing and Doppler boosting that affect rapid variability.  \citet{chan2015b} showed these effects to be important in understanding the broadband variability properties of the supermassive black hole at the center of the Milky Way, Sagittarius  A* (Sgr~A*) and especially for accounting for mm and IR flares that originate close to the event horizon.

Observations of Sgr~A* reveal significant multiwavelength variability (see \citealt{eckart2004,marrone2008,neilsen2013,IAU:9269068,ponti2015,li2015,wang2016}) from the mm and IR to the X-rays.  At high energies, in particular, \citet{neilsen2013} analyzed 3 million seconds of Chandra data dedicated to characterizing both short and long-term X-ray variability of Sgr~A*.  They found that the length of flares varies from a few hundred seconds to 8 ks, with luminosities from $\sim 10^{34}$ erg s$^{-1}$ to $2 \times 10^{35}$ erg s$^{-1}$.  Eckart et al. (2004) carried out simultaneous observations in both the the X-ray and near IR, and found that X-ray flares always have a coincident IR flare. In contrast there are numerous IR flares without X-ray counterparts.  These flares can offer unique insight into the particle energetics of the accretion flow; highly energetic stochastic flares point towards mechanisms such as shocks and magnetic reconnection, which are capable of quickly producing large numbers of high energy non-thermal particles which significantly alter the observational signatures of the accretion flow.  

In this paper, we investigate the effect of incorporating the emission from non-thermal electrons in GRMHD simulations of Sgr~A*.  We consider two injection models throughout the flow for a power-law population of electrons.  In one, the non-thermal electrons simply track the thermal energy throughout the flow. In the second, non-thermal electrons are injected only into regions of possible magnetic reconnection, characterized by a low plasma $\beta$, where
\begin{equation}\label{beta}
\beta \equiv \frac{P_{\rm gas}}{P_{\rm magnetic}}.
\end{equation}
We find that X-ray variability is a generic result of localizing non-thermal electrons to highly magnetized regions and that the timescales for such events are comparable to those of the observed X-ray flares.  These results hold when we consider the cooling and advection of the non-thermal electrons in the flow.

\section{GRMHD Models}

\citet{chan2015a} performed a large study of the broadband, time-dependent emission properties of Sgr~A*.  In these studies, they employed the GRMHD code HARM \citep{gammie2003,narayan2012} in conjunction with the efficient radiative transfer algorithm GRay \citep{chan2013} and varied the BH spin, density normalization, observer inclination, initial magnetic field configuration, and electron thermodynamic prescription.  Among the models in this large parameter space, they chose five that fit the steady-state broadband spectrum of Sgr~A* and its observed 1.3~mm image size.

\citet{chan2015b} then studied the variability properties of the five best fit models at four different frequencies: in the radio at  $10^{10}$ Hz and 1.3~mm, in the infrared at 2.17~microns, as well as in the X-rays at 4.3~keV.  In particular, two models from the study, with a black hole spin of $a=0.7$ and $a=0.9$, standard and normal evolution (SANE) initial magnetic field configuration, and constant electron temperature in the funnel (hereafter referred to as models A and B), showed persistent variability as well as rapid flaring events in the IR and at 1.3 mm, consistent with observations.  \citet{chan2015b} identified these variations as being caused by the dynamic nature of magnetic flux tubes in the flow combined with gravitational lensing that occurs when a flux tube crosses a caustic behind the black hole.  None of the models, however, reproduced any X-ray variability: the X-ray lightcurves were extremely smooth, lacking any notable features from short-timescale variability to longer flaring events.  This is expected given that the X-rays are produced by thermal Bremsstrahlung emission over the entire simulation volume but also indicates that these GRMHD models are missing the physics that causes the rapid variation in the X-ray flux.  

The models in the earlier study use the ideal MHD approximation, where magnetic field dissipation and particle acceleration is not explicitly modeled.  For the remainder of this paper, we use model B from \citet{chan2015a} and incorporate a population of high energy power-law electrons in the postprocessing radiative transfer calculations, which may be accelerated from magnetic reconnection, as described in the following section.

\begin{figure*}
\includegraphics[width =0.5\textwidth]{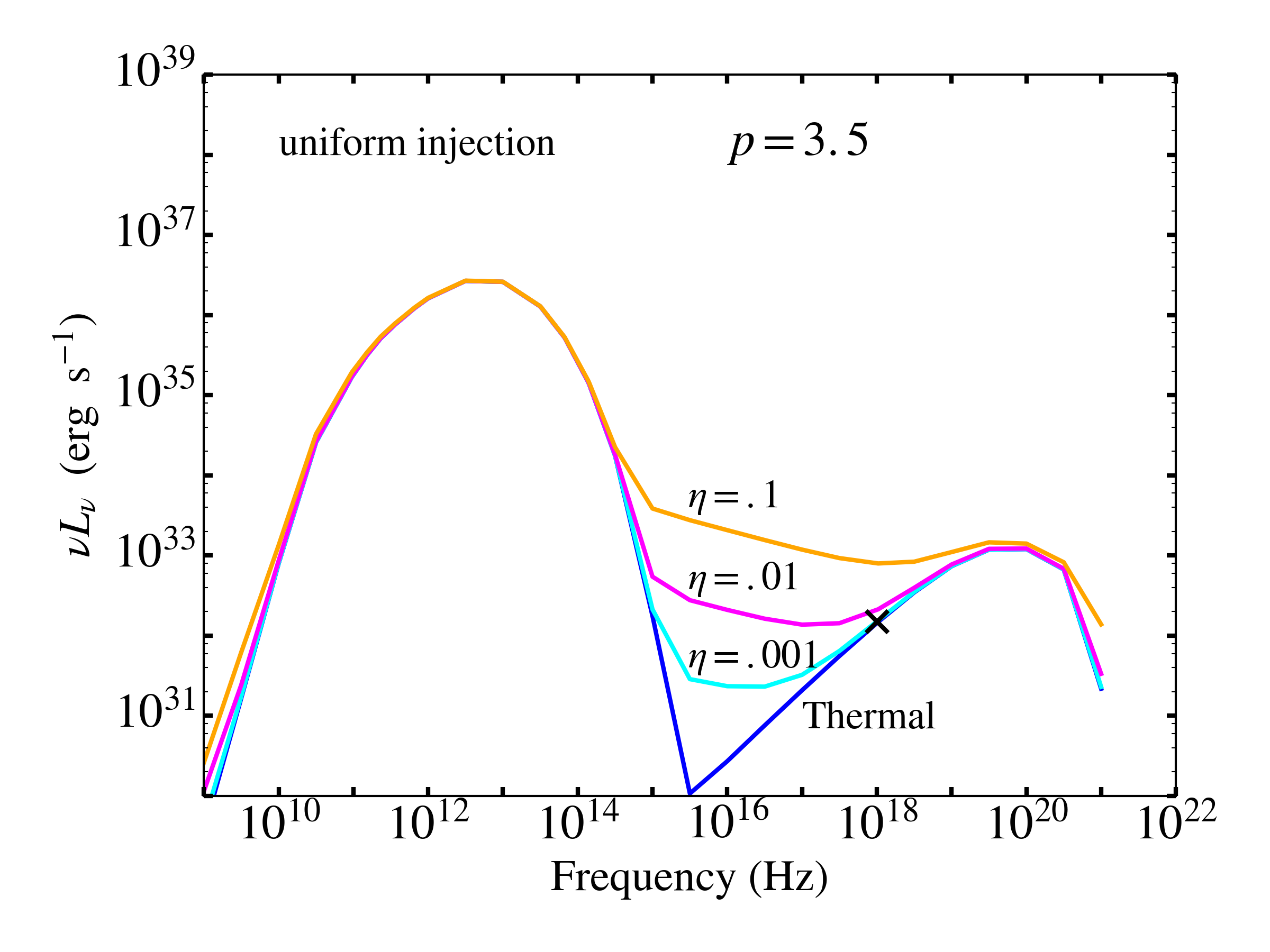}\label{uniform}
\includegraphics[width=.5\textwidth]{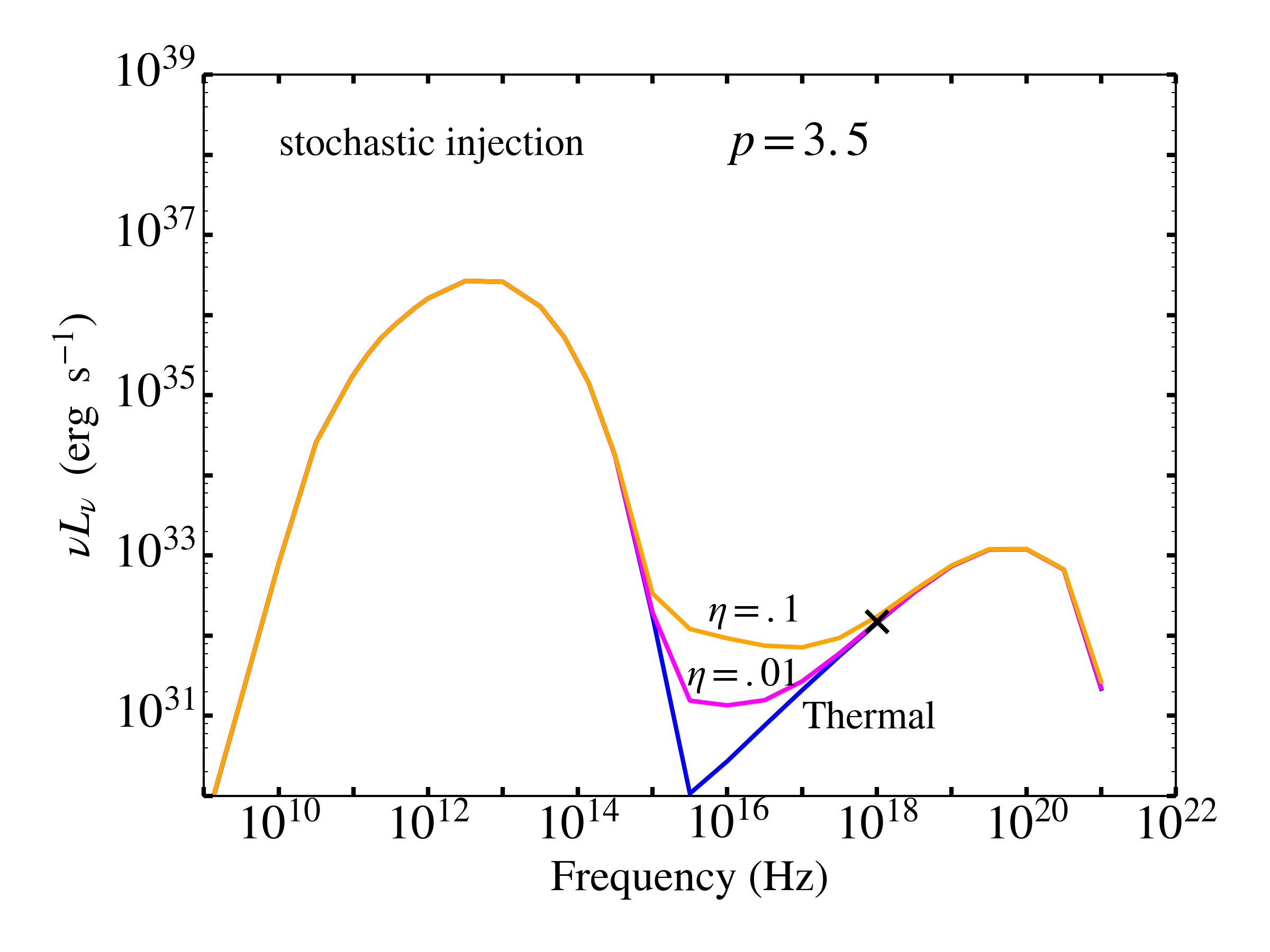}\label{nonuniform}
\caption{\emph{Left:} Spectra computed  for quiescent (i.e., non-flaring) times for various values of the energy fraction of non-thermal electrons, $\eta$, with fixed power-law index, $p=3.5$, as well as for the purely thermal model.  In this configuration where the non-thermals only follow the thermal energy, the observed quiescent X-ray flux at $10^{18}$ Hz (depicted with an X) is exceeded even for moderate values of $\eta$.
\emph{Right:}  In this configuration, non-thermal electrons are injected in regions below $\beta=0.2$.  Localizing the non-thermal electrons to highly magnetized regions, where they are more likely to be accelerated, allows for significantly higher values of $\eta$ while still accommodating the observed quiescent X-ray flux at $10^{18}$ Hz.}
\end{figure*}
\section{Incorporating Non-Thermal Electrons}
A population of particles in the accretion flow will evolve according to the continuity equation
(written here in flat spacetime for simplicity) 
\begin{equation}
\frac{\partial n_\gamma}{\partial t} + \vec{v}\cdot \vec{\nabla} n_\gamma
+\left(\vec{\nabla}\cdot\vec{v}\right) n_\gamma =\left.\frac{dn_\gamma}{dt}\right\vert_{\rm inj}+
\dot{\gamma}\left.\frac{dn_\gamma}{d\gamma}\right\vert_{\rm cool}
\label{continuity}
\end{equation}
where $n_\gamma$ is the number density of electrons with Lorentz factor $\gamma$ and $\vec{v}$
is their bulk velocity. The first term describes the evolution of the electron energy distribution, the second
term represents the advection of particles with the flow, the third term describes the effect of the
convergence/divergence of the flow, the fourth term describes the rate of injection of particles from
acceleration processes such as magnetic reconnection or shocks, and the final term accounts
for the cooling of particles, with $\dot{\gamma}$ being the radiative cooling rate.

The rest of this paper focuses on understanding each of these different terms and their
relative contributions, which depend strongly on the local properties of the accretion flow. Because
both the fluid velocity and its density have similar power-law dependences on radius, the second
and third terms in the above equation have comparable magnitudes.  For this reason, we will
not consider further the terms that describes the convergence/divergence of the flow.
\subsection{Injection of Non-Thermal Electrons}
We consider two configurations for the injection of non-thermal electrons in the accretion disk, i.e., the fourth term in equation (\ref{continuity}).  The first model is based on the assumption that some fraction of the electron heating will continuously go into the acceleration of a non-thermal population and this fraction is independent of conditions in the flow.  This results in a steady and uniform injection of non-thermal electrons, where the population of high energy particles simply follows the thermal energy in the system.  We refer to this scenario as the ``uniform" or ``steady-state" distribution.  

The second configuration is physically motivated by PIC simulations of magnetic reconnection, the process through which opposing magnetic fields are pushed together and dissipate.  Magnetic reconnection rapidly injects energy into a small region of plasma, which has been shown to generate a large population of high energy power-law electrons in regions of low $\beta$ (Sironi \& Spitkovsky 2014).  Motivated by these PIC simulations, we pick a threshold value, $\beta_{t}=0.2$, typical of the funnel or highly magnetized flux tubes, below which we inject non-thermal electrons (note that lower $\beta$ indicates higher magnetic pressure).  In this configuration, the dynamic nature of the flux tubes causes the injection to be more variable.  As such, we refer to this model as ``stochastic'' or ``nonuniform''.  


\subsection{Radiative Cooling of Power-Law Electrons}
The densities in the flow are sufficiently low that thermalization via Coulomb collisions is negligible, while the relatively high magnetic fields in the system imply that synchrotron radiation is the dominant cooling process. In order to discuss the role of synchrotron radiation in the cooling of electrons, we first define our electron energy distribution.

We consider a power-law distribution of electrons motivated by the results of magnetic reconnection models 
\begin{equation} \label{power law}
n_{\gamma} d\gamma = C \gamma^{-p}d\gamma,
\end{equation}
where $p$ is the power law index, and $C$ is a normalization constant.  This distribution of non-thermal electrons will radiate predominantly via synchrotron due to the presence of magnetic fields.  The synchrotron power per unit volume per unit frequency emitted by this distribution is given by (e.g., \citealt{rybicki1979}):
\begin{equation} \label{emissivity}
\begin{aligned}
P_{\rm tot}\left(\omega\right) =  \frac{\sqrt{3}q^{3}CB\sin{\alpha}}{4\pi m_{e}c^{2}\left(p+1\right)}\Gamma\left(\frac{p}{4}+\frac{19}{12}\right)\times \\ \Gamma\left(\frac{p}{4}-\frac{1}{12}\right)\left(\frac{m_{e}c\omega}{3qB\sin{\alpha}}\right)^{-\left(p-1\right)/2},
\end{aligned}
\end{equation}
where $q$ and $m_{e}$ are the charge and mass of an electron, $C$ is the normalization constant from equation (\ref{power law}), $B$ is the magnetic field strength, $\alpha$ is the pitch angle between the electron velocity and magnetic field, and $\Gamma$ is the gamma function. 
The non-thermal electrons will also add to the opacity of the system as
\begin{equation} \label{opacity}
\begin{aligned}
\alpha_{\nu} = \frac{\sqrt{3}q^{3}}{8\pi m_{e}}\left(\frac{3q}{2\pi m_{e}^{3}c^{5}}\right)^{p/2}C\left(B\sin{\alpha}\right)^{\left(p+2\right)/2}\times \\ \left(m_{e}c^{2}\right)^{p-1}\Gamma\left(\frac{3p+2}{12}\right)\Gamma \left(\frac{3p+22}{12}\right)\nu^{-\left(p+4\right)/2}.
\end{aligned}
\end{equation}
Assuming an isotropic velocity distribution, we average equations (\ref{emissivity}) and (\ref{opacity}) over the pitch angle in our calculations.

We assign a fraction of the total thermal energy to this power-law distribution, which then radiates and absorbs photons according to equations (\ref{emissivity}) and (\ref{opacity}).  The energy density of thermal electrons at temperature $\theta_{e}$ is (e.g., \citealt{chandrasekhar1939})
\begin{equation} \label{thermal energy}
u_{th} = a\left(\theta_{e}\right)N_{th}m_{e}c^{2}\theta_{e},
\end{equation}
where $N_{th}$ is the number density of thermal electrons, $a\left(\theta_{e}\right)$ is given by
\begin{equation}
a\left(\theta_{e}\right) \equiv \frac{1}{\theta_{e}}\left[\frac{3K_{3}\left(1/\theta_{e}\right)+K_{1}\left(1/\theta_{e}\right)}{4K_{2}\left(1/\theta_{e}\right)}-1\right],
\end{equation}
and $K_n$ are modified Bessel functions of order $n$.  The quantity $a\left(\theta_{e}\right)$ varies from 3/2 to 3, corresponding to a non-relativistic and fully relativistic electron gas, respectively.  In order to aid computation, we use an approximate form for $a\left(\theta_{e}\right)$ given by \citet{gammie1998}, which has less than 2\% error for all temperatures:
\begin{equation}
a\left(\theta_{e}\right) = \frac{6 + 15 \theta_{e}}{4 + 5\theta_{e}}.
\end{equation}
We now introduce a free parameter, $\eta$, which describes the fraction of thermal energy assigned to a power-law distribution.  The non-thermal energy density in a given cell is, therefore, $u_{pl} = \eta u_{th}$.
The quantity $C$, in equations (\ref{emissivity}) and (\ref{opacity}) is related to $\eta$ by
\begin{equation} \label{eta calculation}
\eta u_{th} = \int_{\gamma_{1}}^{\gamma_{2}}C\left(\gamma mc^{2}\right)\gamma^{-p}d\gamma
\end{equation}
We choose $\gamma_{1}=1$ and set $\gamma_{2}$ to be very large so that
\begin{equation}
C \approx \eta a\left(\theta \right)N_{th}\theta_{e}\left(p-2\right) 
\end{equation}

With this set up, we can calculate the necessary quantities to perform the radiative transfer calculation while accounting for a population of non-thermal electrons described by the quantities $p$ and $\eta$.  In \textsection 4, we discuss the results of this calculation for our two different models, which are the uniform steady state injection and the $\beta$-dependent, nonuniform, stochastic injection.  In \textsection 6 we more rigorously discuss the advection and synchrotron cooling timescales to complete our analysis of equation (\ref{continuity}). 

\begin{figure}[t]
\includegraphics[width =0.5\textwidth]{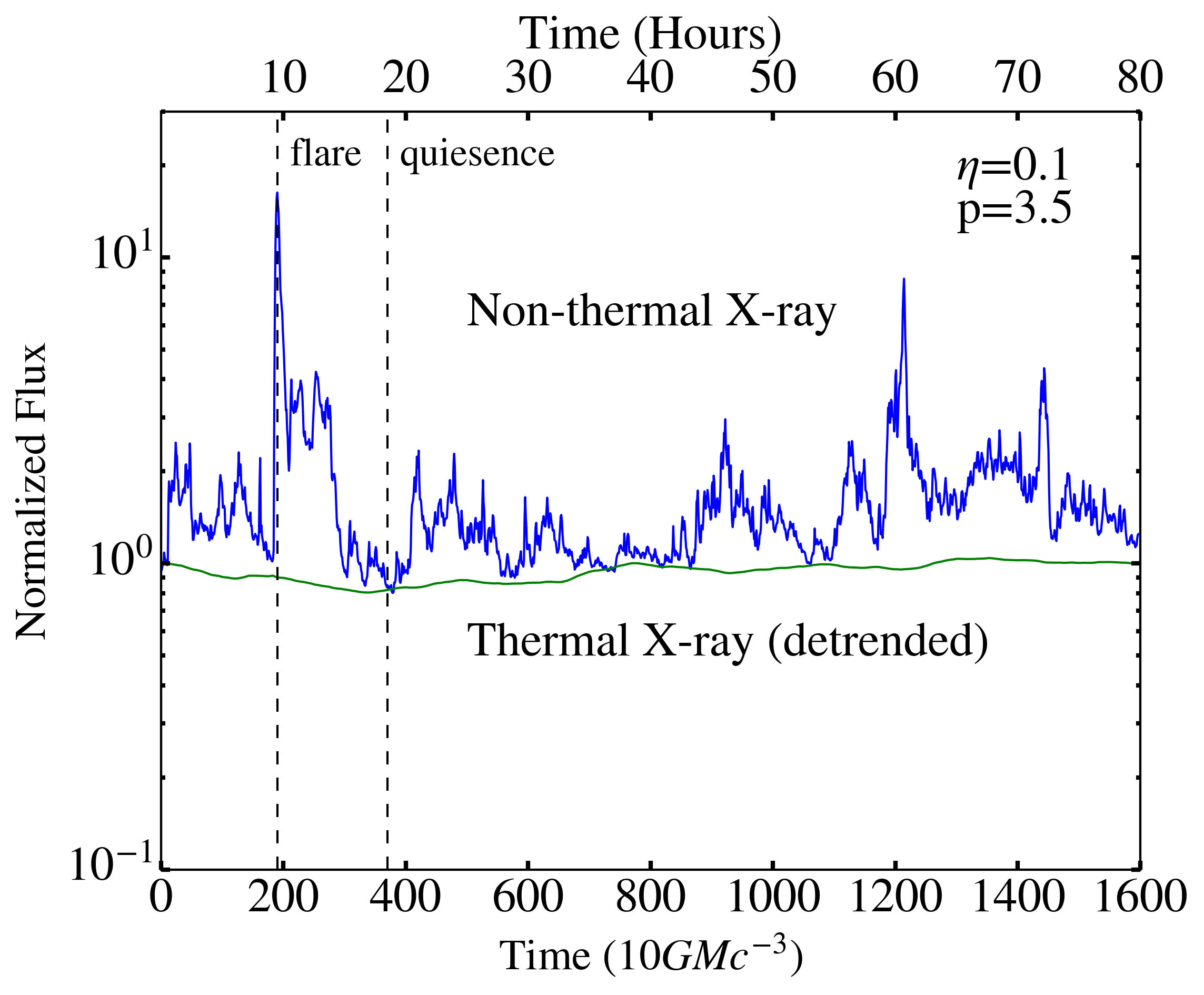}
\caption{Thermal and non-thermal X-ray lightcurves.  The injection of non-thermal electrons into highly magnetized regions naturally produces significant variability due to the dynamic nature of magnetic fields in the accretion flow.}
\label{therm_vs_nontherm}
\end{figure}

\section{Quiescent Constraints}
In \citet{chan2015a}, purely thermal models were fit to a number of observed quantities, including to the quiescent X-ray flux at $10^{18}$ Hz ($\equiv$ 4.1 keV), which originates predominantly from the extended halo of gas emitting via Bremsstrahlung.  The thermal models were calibrated to reproduce the appropriate observed time-averaged X-ray flux for the size of the simulation, corresponding to 10\% of the total observed flux.  As such, when we include the non-thermal electrons, a natural requirement is that the quiescent X-ray flux must not change significantly compared to the thermal model. 

\begin{figure*}
\includegraphics[width =0.5\textwidth]{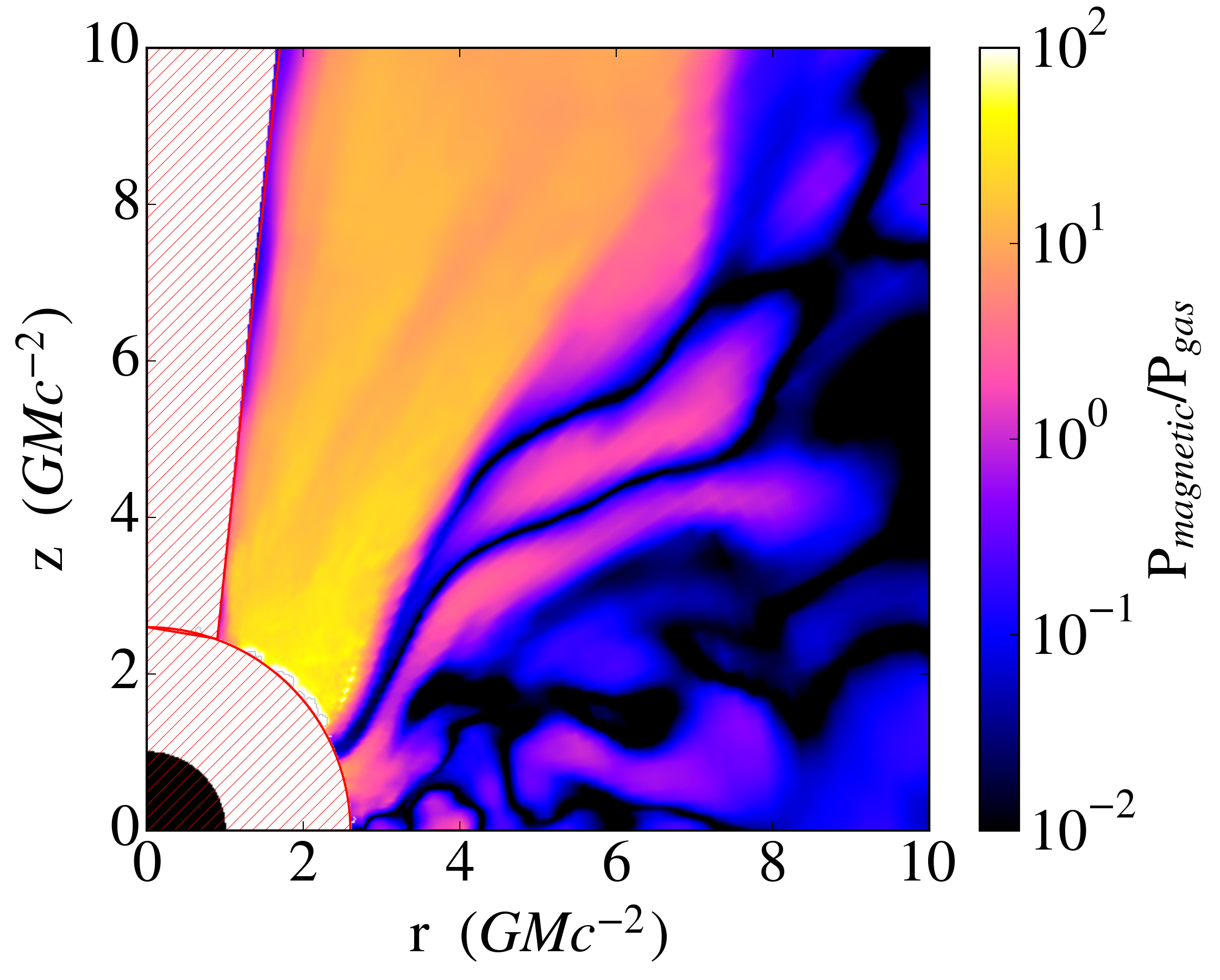}\label{quiescent_beta}
\includegraphics[width=.5\textwidth]{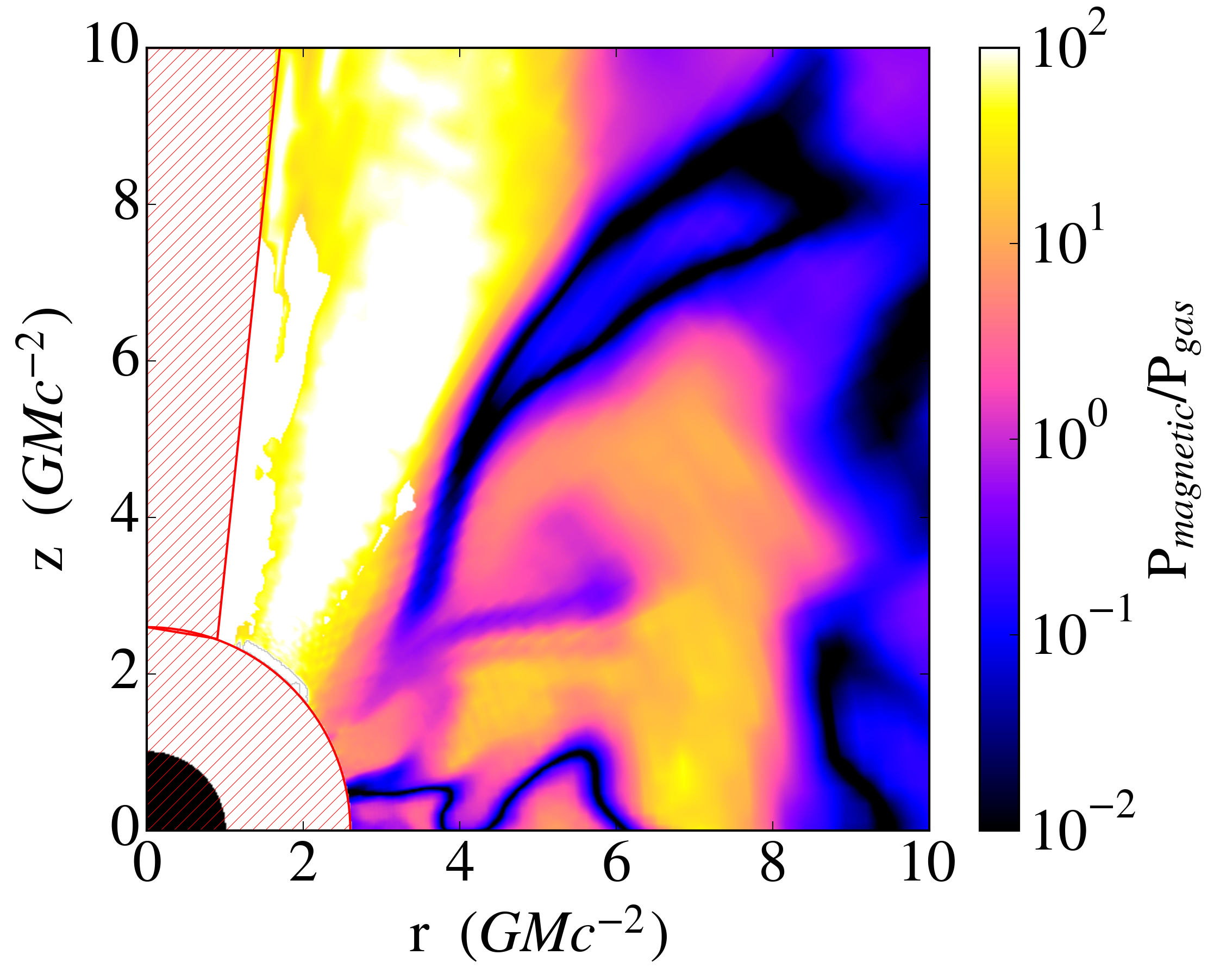}\label{flare_beta}
\caption{
\emph{Left:} Map of the ratio of the magnetic to gas pressure during a quiescent state in the simulation.  Cells near the pole and within the ISCO at $\sim$ 2.4 $GMc^{-2}$ are excised due to numerical artifacts often occurring within these regions.  The black quarter-circle at the origin is the event horizon of the black hole.  \emph{Right:}  A large flux tube is present in the accretion flow during this flare, with high magnetization, resulting in a high ratio of pressures throughout a large portion of the disk.}
\end{figure*}

We show in the left panel of Figure 1 the results of the uniform steady-state model with a power-law index of 3.5, which produces a spectral index within the bounds of observational constraints (\citealt{barriere2014,porquet2008}) and is motivated physically by PIC simulations of magnetic reconnection for magnetizations on the order of the regions we are considering \citep{sironi2014}.  It is evident that, in order to match the X-ray flux, the fraction of energy in non-thermal electrons should be quite small, constrained to values of the order of 0.001.  This implies that there cannot be a very large population of non-thermal electrons existing everywhere throughout the flow at any given time; even a relatively small number of these high energy electrons will result in too large of an X-ray flux if they are distributed throughout the entire simulation region.   

In the right panel of Figure 1, we show the spectrum from the nonuniform stochastic model, where non-thermal electrons are localized to low $\beta$ regions, again with a power-law index of 3.5.
Because of this localization, it is possible to accommodate higher fractions of non-thermal electrons while still matching the quiescent thermal spectrum.  In this model, it is possible to inject almost 10\% of the total thermal energy into a non-thermal electron distribution within the magnetized regions.

\section{X-ray Variability: Stochastic Injection}
Apart from providing a more natural match to quiescent-state constraints, another interesting result of using the $\beta$-dependent description of non-thermal electron injection is that it produces significant X-ray variability.  Perhaps this is not surprising, since  the non-thermal electrons will trace magnetic flux tubes, which are dynamic structures, constantly being formed, sheared, and moving throughout the flow.  If one of these flux tubes crosses a caustic behind the black hole, it will result in an additional amplification of the flux, and since these tubes are emitting primarily non-thermal synchrotron radiation in the X-rays, they will cause X-ray flares.  

We explore this variability in Figure \ref{therm_vs_nontherm}, where we show the effect of stochastic injection of non-thermal electrons and compare it to a purely thermal model.  In the non-thermal lightcurve, we see both persistent variability as well as 4 large flares during the $\sim$80 hours of simulation.  In the largest flares, the flux increases by a factor of $\sim$10 compared to quiescence.  The magnetically dominated regions responsible for these flares live for about 5000 seconds, which sets the timescale of the flares in this figure.  There is indeed a stark contrast between this result, which takes into account acceleration in low $\beta$ regions, and the purely thermal model, which shows no variability.

We now investigate the properties of the magnetic structures in the innermost regions of the accretion flow to further pinpoint the localization and time evolution of the flares.  Figure 3 shows the ratio of the magnetic to the gas pressure throughout the inner flow during a quiescent state and during the strongest flare from the simulation.  We see that this flare is caused by a large magnetic flux region developing in the flow with $\beta < \beta_{t}$.  Due to the large spatial extent of this tube, many non-thermal electrons are injected, causing a sudden increase in the X-ray flux.  In contrast, during quiescence, the only region with a significant number of non-thermal electrons is in the funnel, which typically has a fairly uniform and strong magnetic field.  This only contributes a small flux and results in low level variability.  


\begin{figure}[t]
\includegraphics[width =0.5\textwidth]{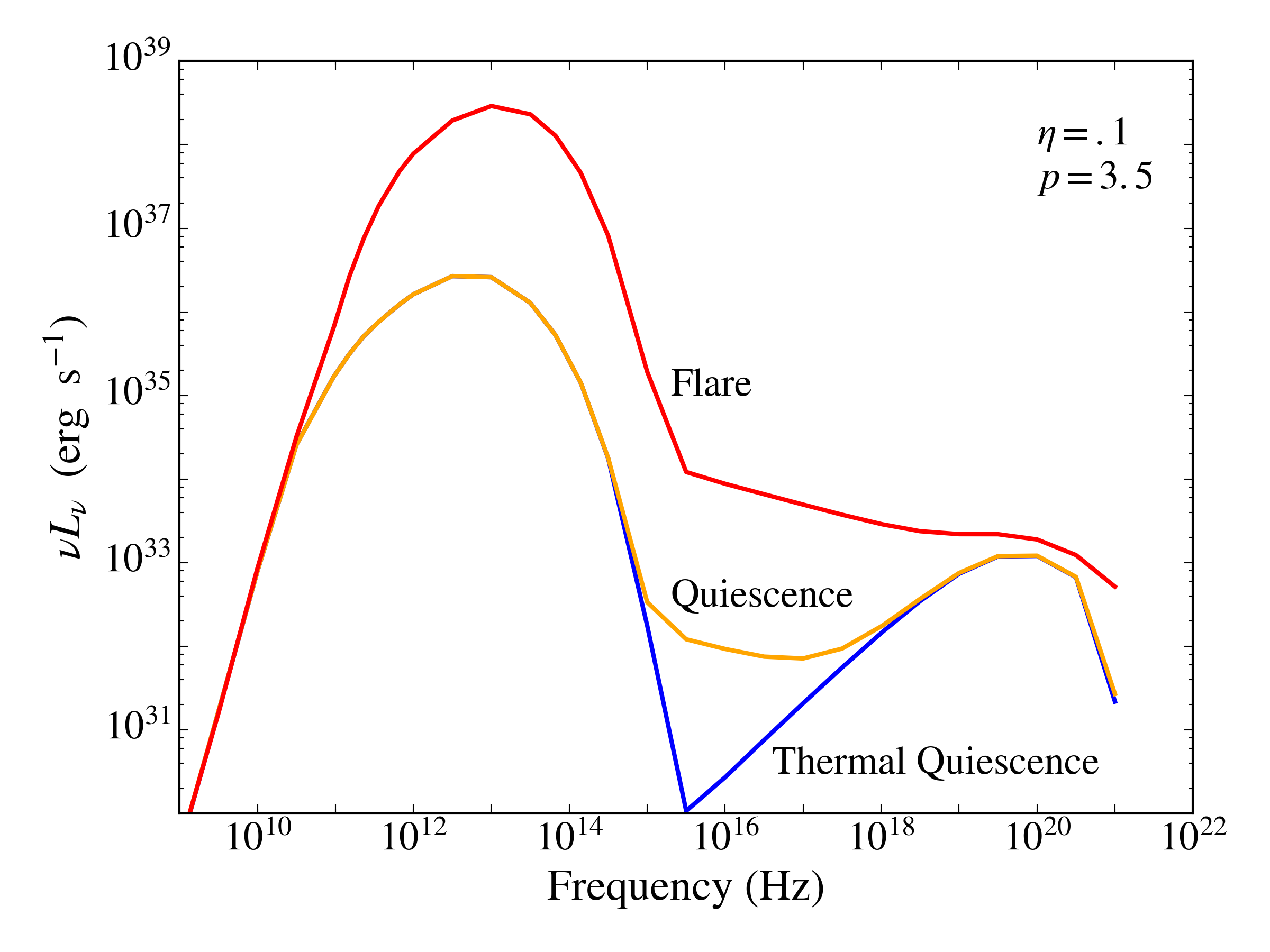}
\caption{Spectra of the flaring and quiescent states depicted in Figure 3 in red and orange, respectively.  The purely thermal quiescent spectrum is shown for reference in blue.  }
\label{spectra}
\end{figure}
Figure \ref{spectra} depicts the spectra of the flaring and quiescent states from the simulation.  During quiescence, the non-thermal emission is not especially prominent; its nature is largely obscured by the thermal emission dominating at most wavelengths.  During the flare, however, the power-law nature of the non-thermal emission becomes more evident.

In Figure 5, we show the X-ray images of our model during flaring and quiescent states.  The images show the relative contribution to the overall flux from various parts of the accretion flow.  During quiescence, we find that while there is some contribution to the X-ray flux from a small population of non-thermal particles in the funnel, the extended Bremsstrahlung emission accounts for the majority of the total (i.e., integrated over the entire image) observed flux.  During a flare, the emission is heavily dominated by non-thermal electrons in the inner accretion flow, rendering the Bremsstrahlung flux negligible.  The difference in the localization of the X-rays between the non-thermal and thermal models is responsible for their different variability properties.

\begin{figure*}
\includegraphics[width =0.5\textwidth]{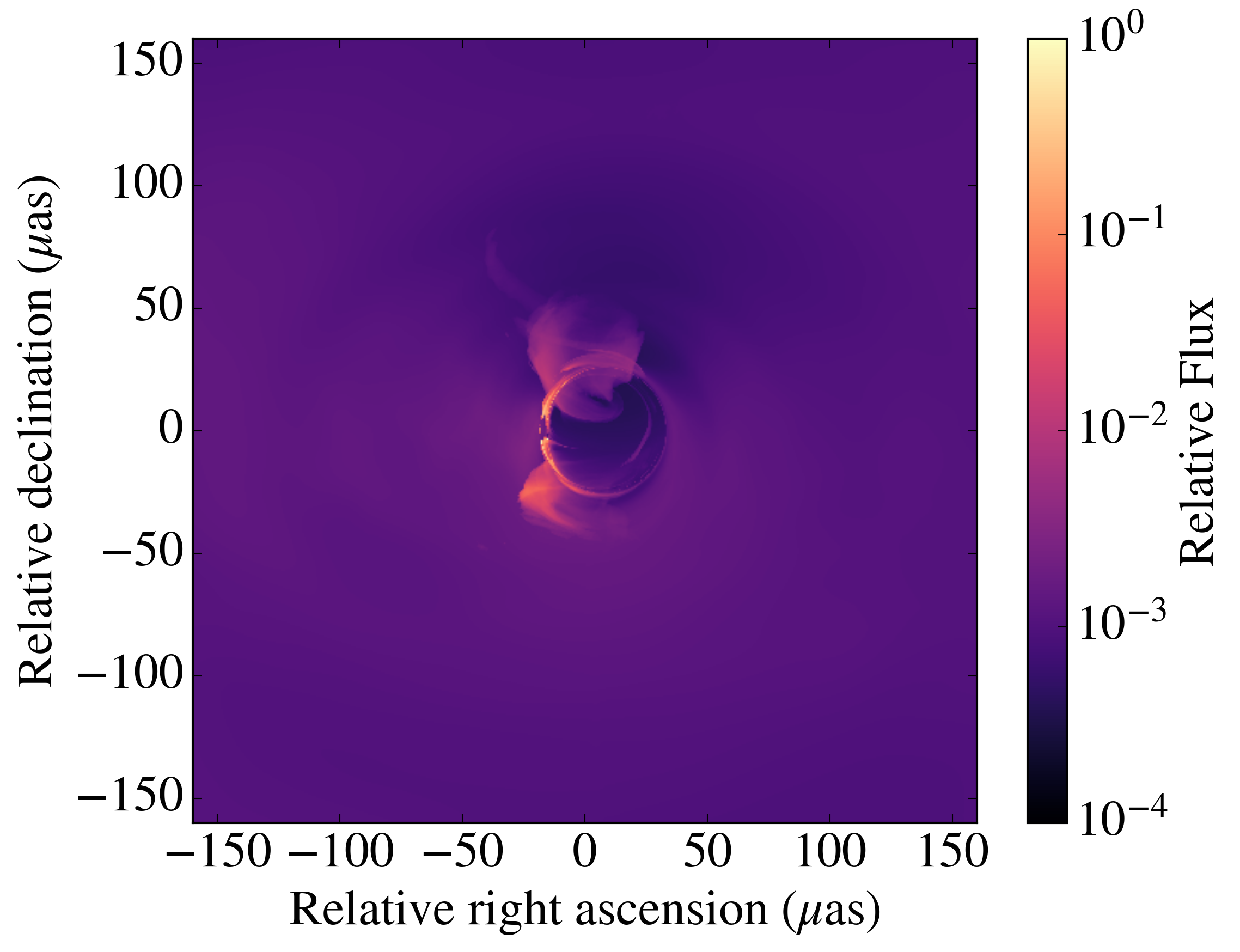}\label{quiescent_image}
\includegraphics[width=.5\textwidth]{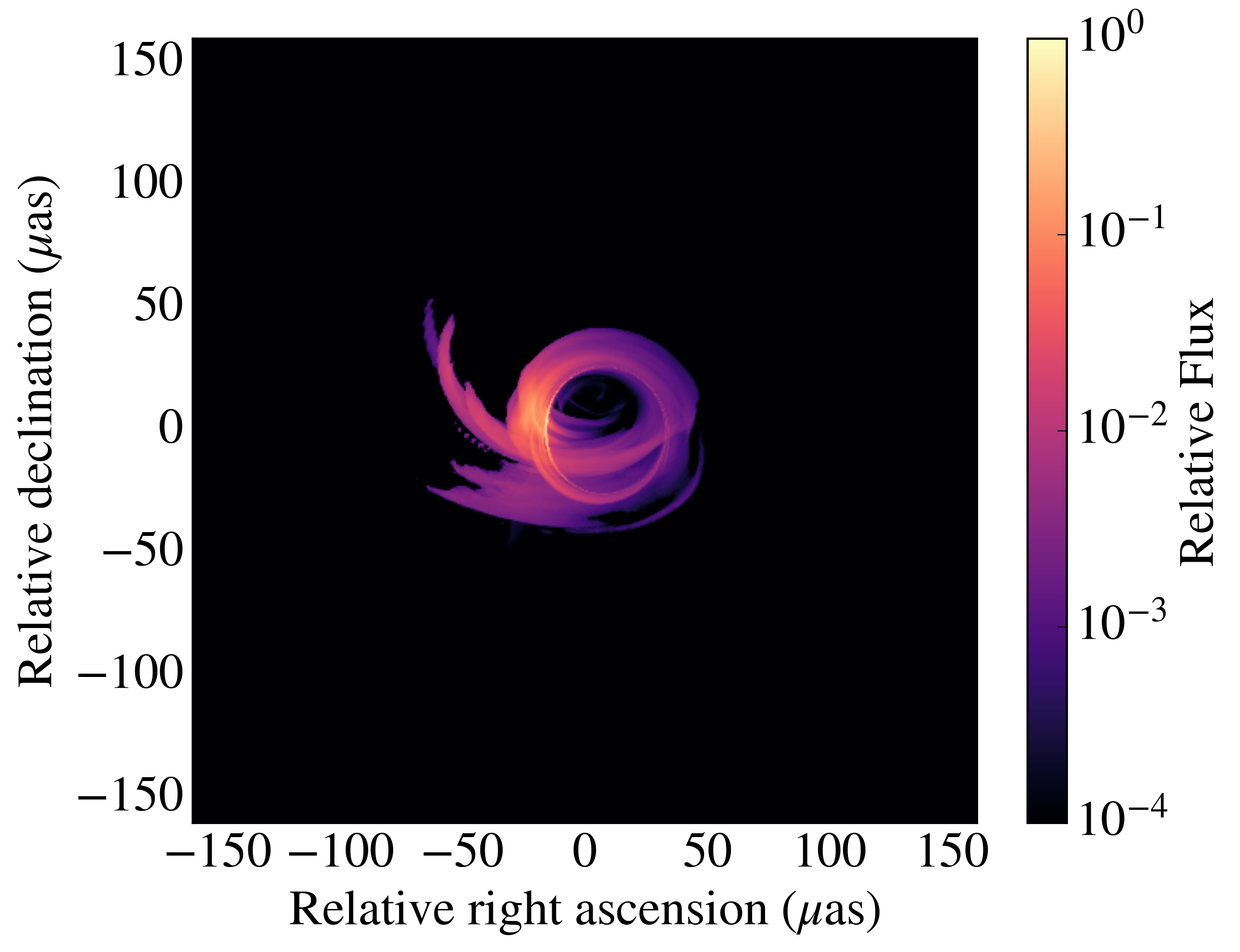}\label{flare_image}
\caption{
\emph{Left:} Simulated image of quiescent X-ray (4.1 keV) emission.  Fluxes are normalized to the maximum pixel value.  Some structure is visible in the innermost regions of the image, where strong magnetic fields in the funnel close to the event horizon have associated non-thermal particles, and hence strong X-ray emission.  We see that the extended Brehmsstrahlung emission comprises a significant fraction of the total flux during quiescence.
\emph{Right:}  During the flare, emission is heavily dominated by the innermost part of the accretion flow; the relative contribution from the halo of Bremsstrahlung emission is negligible during flares.}
\end{figure*}

\section{Cooling and Advection timescales}
If the production of non-thermal electrons in the flow are indeed stochastic and localized, their lifetime will be determined by the cooling and advection timescales relative to the injection timescale.  In this section, we consider the cooling and advective terms in equation (\ref{continuity}) and compare the typical advective timescales for matter to be drawn through the event horizon to the synchrotron cooling times.  These timescales will give us a rough estimate of how long a flare will last.  

\begin{figure*}
\includegraphics[width =0.5\textwidth]{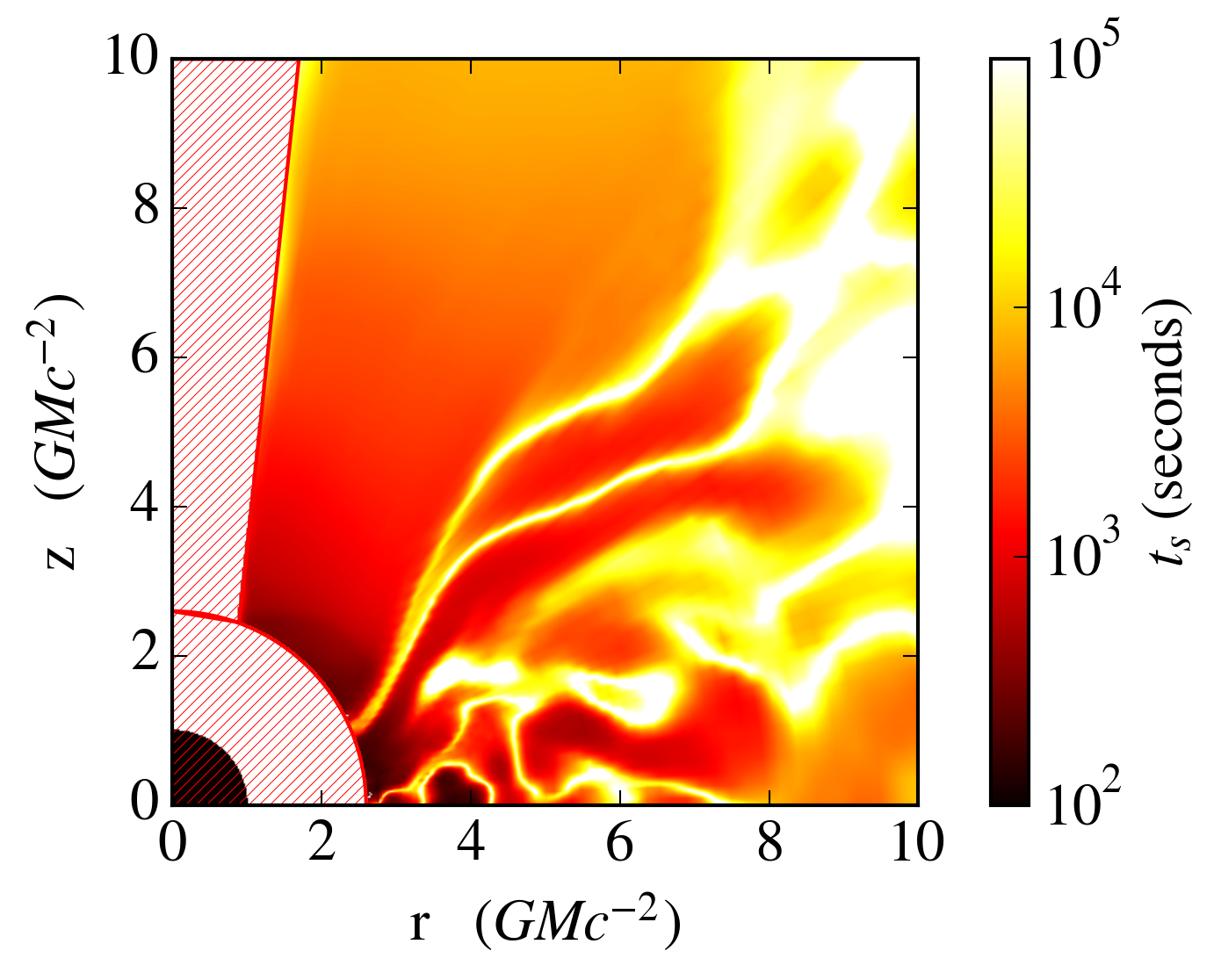}\label{synch_timescale}
\includegraphics[width=.5\textwidth]{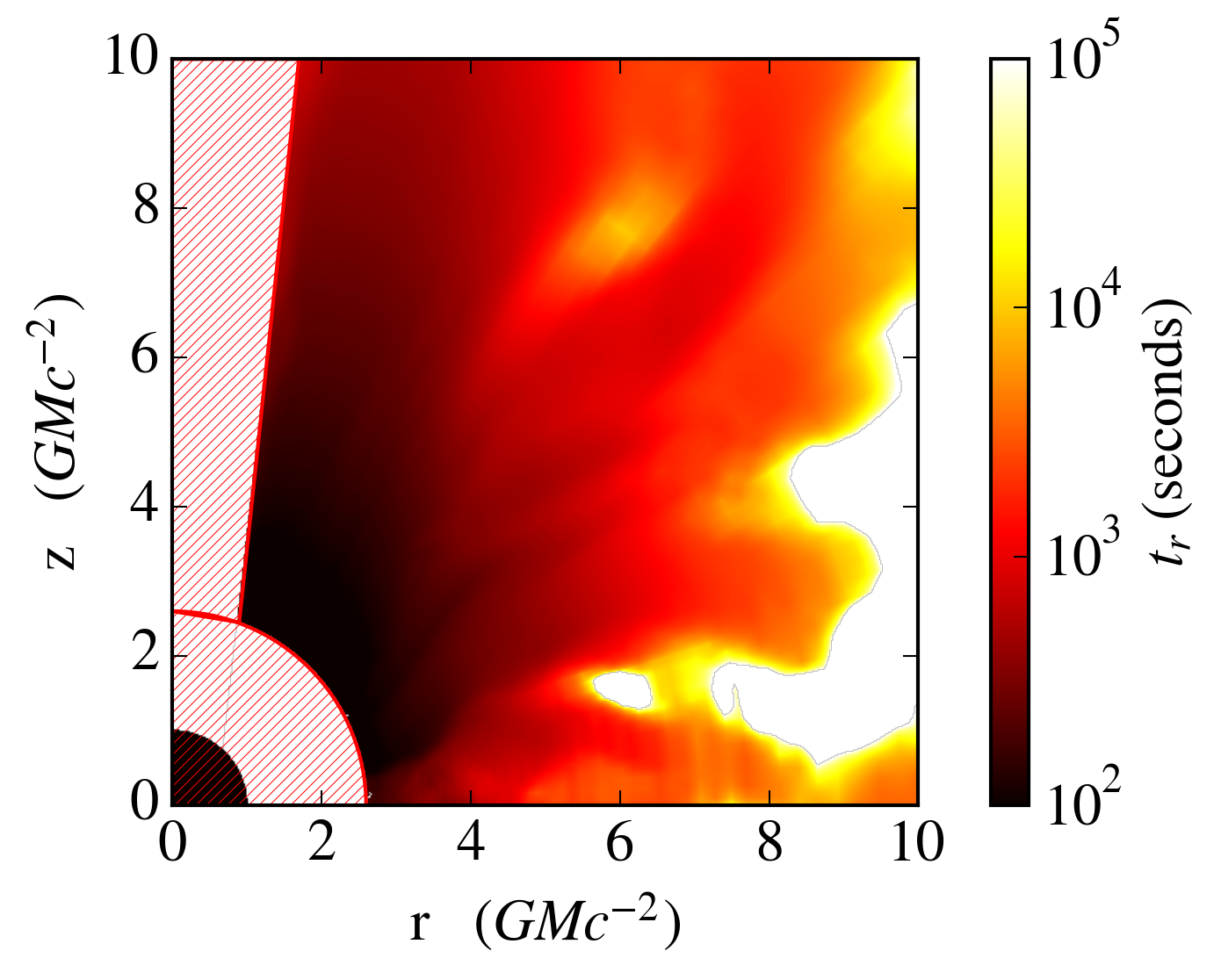}\label{velocity_profile}
\caption{
\emph{Left:} Synchrotron cooling timescale $t_{s}$ in the inner accretion flow.  Cooling timescales in the funnel are of the order of a few thousand seconds, while cooling timescales in the plane of the disc vary greatly due to the nonuniformity of the magnetic field.
\emph{Right:}  Radial advection timescale, $t_{r}$, i.e., an estimate of how long it will take for the matter at a given point to pass through the event horizon.  Matter in the funnel is quickly advected on the order of hundreds of seconds, while matter in the plane of the disk, $z = 0$, is centrifugally supported and has radial advection timescales of the order of hundreds to thousands of seconds.}
\end{figure*}
The synchrotron cooling time is given by (e.g., \citealt{rybicki1979})
\begin{equation}
t_{s} = \frac{6 \pi m_{e} c}{\sigma_{t} \beta_{e}^{2} \gamma B^{2}},
\end{equation}
where $\sigma_{t}$ is the Thompson cross section, $m_{e}$ is the electron mass, $c$ is the speed of light, $\beta_{e}$ is the particle velocity as a fraction of the speed of light, $\gamma$ is the Lorentz factor, and $B$ is the magnetic field strength.

We calculate the synchrotron cooling time for the innermost region in the simulation, considering fairly relativistic electrons with $\gamma = 40$ (a Lorentz factor relevant to X-ray production that is easily generated by relativistic reconnection) and show the results in Figure \ref{synch_timescale}.

The cooling timescale in the funnel is of the order of thousands of seconds, while it varies greatly in the disk due to the highly variable magnetic field.  The magnetized flux tubes have cooling times of the order of hundreds to thousands of seconds, depending on the magnetization of the region, while the filamentary regions in between them have extremely long timescales, corresponding to weak magnetic fields.  Most importantly, we see that the synchrotron cooling timescales in the regions where we expect there to be non-thermal electrons (i.e., low $\beta$ regions) are comparable to the observed flare durations of a few hundred seconds to 8000 seconds, reported in \citet{neilsen2013}.  

\begin{figure*}
\includegraphics[width =0.5\textwidth]{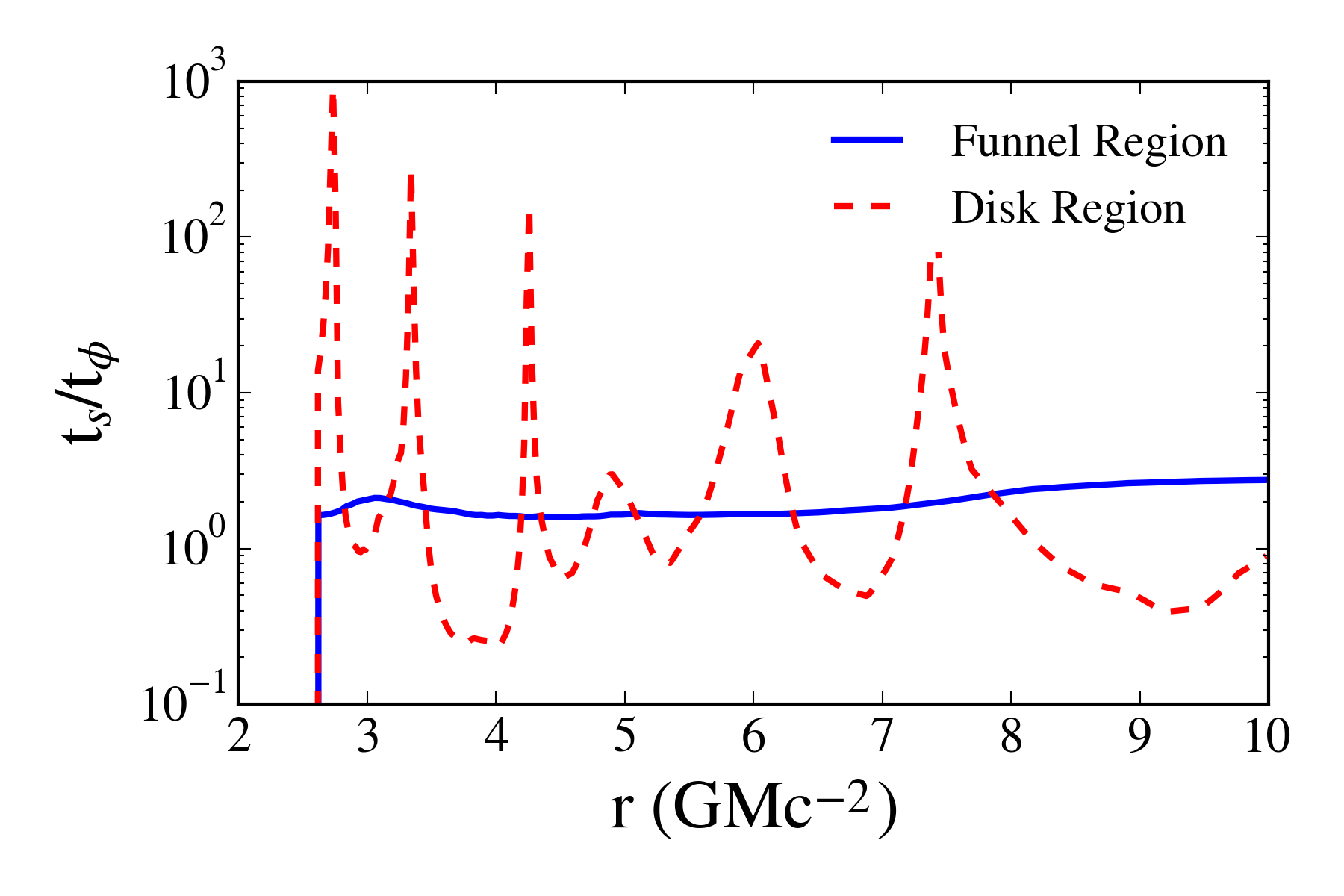}\label{funnel_disk_synch}
\includegraphics[width=.5\textwidth]{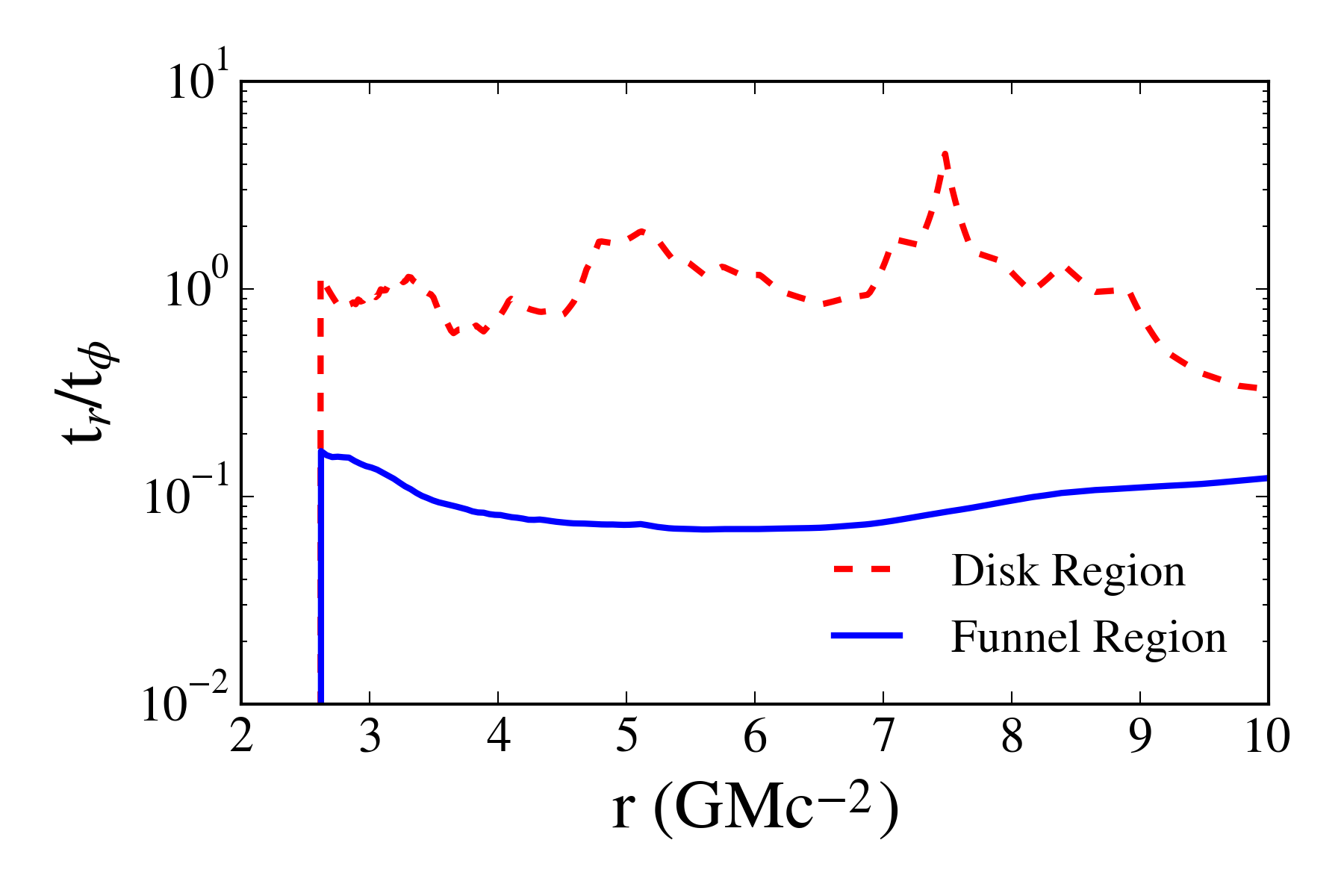}\label{funnel_disk_advec}
\caption{
\emph{Left:} Synchrotron cooling timescales in terms of azimuthal dynamical time.  The synchrotron cooling in the funnel is fairly constant due to the uniform magnetic field.  The disk synchrotron is highly varied due to the structure of the magnetic field
\emph{Right:}  Radial advective timescale in terms of azimuthal dynamical time.  The centrifugal support in the disk allows the radial time to be comparable to the azimuthal time, while matter in the funnel will last at most for one-tenth of an orbit before being drawn through the event horizon.}
\end{figure*}

Next, we consider the radial advective timescale, which is the time it will take electrons to move radially a distance r in the flow.  We define the advection timescale as $t_{r} \equiv -r/v_{r}$.  In the model we are considering, all the fluid within 10 gravitational radii of the event horizon has an inward (negative) radial velocity, such that $t_{r}$ is always positive, as shown in Figure \ref{velocity_profile}. 

Because radial advection is inward, the timescale associated with this motion is relevant to the estimation of flare duration.  If a population of non-thermal electrons is injected and quickly disappears through the event horizon due to radial advection, then the length of the flare may be dictated by the advection timescale.  If, however, $t_{r} > t_{s}$, then the synchrotron cooling will be primarily responsible for the dissipation of non-thermal energy.

The right panel of Figure 6 shows that advective timescales in the funnel are fairly short: any non-thermal electrons injected into the funnel will quickly pass through the event horizon within hundreds of seconds.  
Centrifugal support in the disk results in longer timescales, allowing for a population of instantaneously injected non-thermal electrons to radiate their energy away over the hundreds to thousands of seconds it takes for synchrotron emission to cool them.
As such, when we interpret observations in the context of a GRMHD model such as model B, we conclude that the flaring events predominantly originate from the disk.  In other words, in order for a flare to last for many thousands of seconds as observed, the non-thermal electrons must be in the disk where they are centrifugally supported and able to exist for these timescales in the flow without disappearing through the event horizon.  

We now consider the advective and synchrotron timescales in terms of the azimuthal dynamical time, $t_{\phi}=2 \pi r / v_{\phi}$.  By doing this, we get a sense of how much a population of electrons will be able to spread azimuthally throughout the flow as they cool.  In Figure~7, we plot the synchrotron and advective cooling times in units of azimuthal dynamical time, taking a slice of the simulation through the disk ($z=0$) shown in Figure~\ref{synch_timescale}, and a slice through the funnel, making a 60 degree angle with the horizontal axis in Figure~6.



We see the same general features we already discussed in terms of the structure and relative magnitude of synchrotron and advective timescales in both the funnel and the disk.  However, we now get more physical insight into the dynamics of the system: we find that, in the highly magnetized flux tubes, corresponding to local minima of the dashed red line in the left panel of Figure 7, the synchrotron cooling is sufficiently fast that a flare will be localized to a fairly small azimuthal region, since t$_{s}$ $<$ t$_{\phi}$.  Similarly, we find that any matter injected into the funnel will be radially advected much more quickly than it is able to spread out azimuthally.

\section{Comparison to Observations}
By interpreting observations of Sgr~A* in the context of our GRMHD simulations, we begin to place constraints on the population of non-thermal electrons in the radiatively inefficient accretion flow and gain insight into their injection mechanisms.    
We employed two configurations for non-thermal electron injection, one in which the non-thermal electrons simply track the thermal energy everywhere throughout the flow and another where the non-thermal electrons are injected solely into regions of high magnetization.  From the first model we are able to place tight constraints on the fraction of steady-state non-thermal electrons that may exist throughout the flow by comparing the simulations to the observed quiescent X-ray flux.  The second model localizes non-thermal electrons to a much smaller region, allowing their local energy density to be much higher than the uniform model (by about 2 orders of magnitude) while still matching the observed quiescent X-ray flux.

\begin{figure}[!h]
\includegraphics[width =0.5\textwidth]{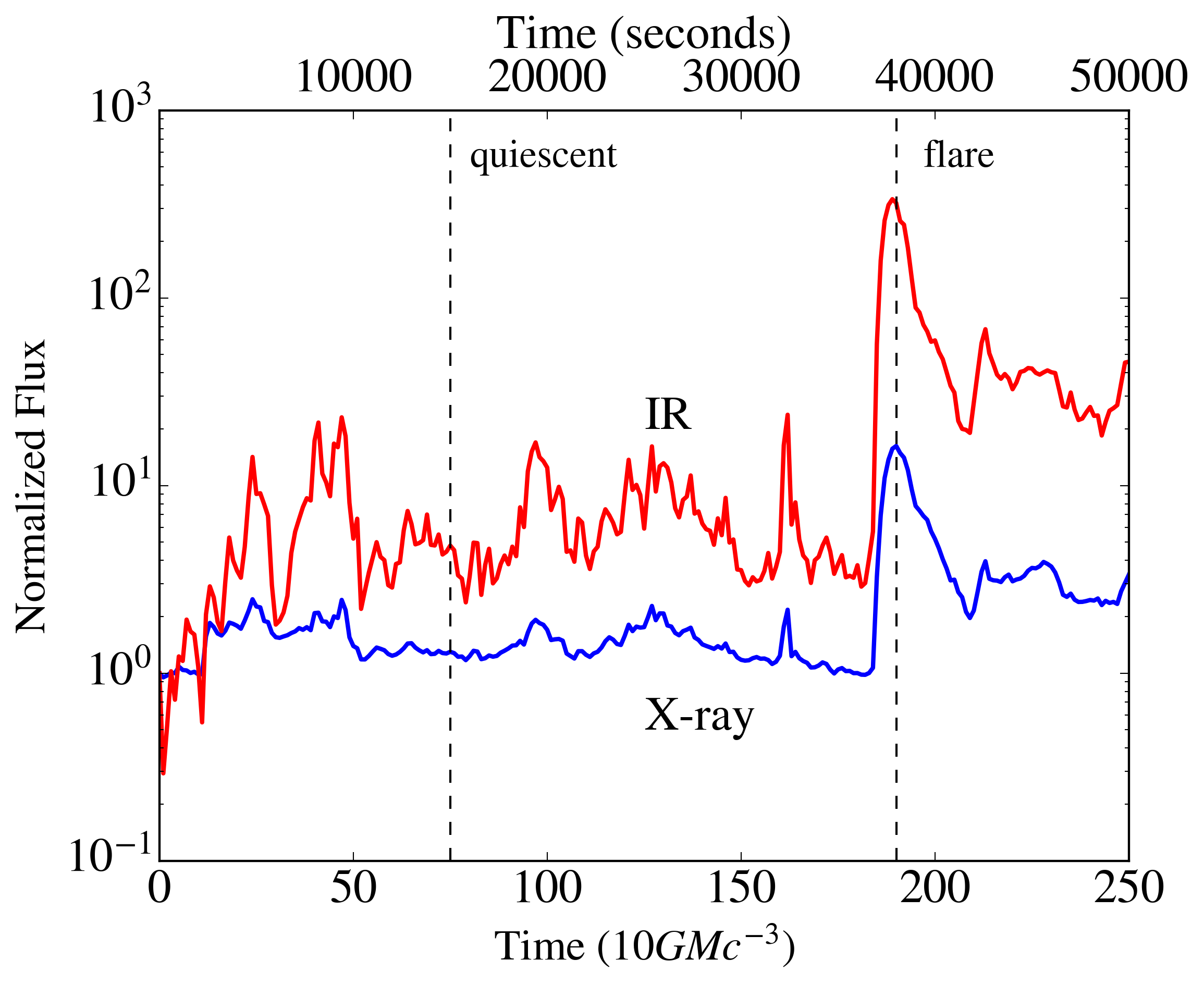}
\caption{IR and X-Ray lightcurves zoomed in on the first 250 timesteps of the simulation.  Each curve is normalized to a fiducial quiescent flux.  Note the strong and rapid variation in the IR flux and moderate variability in the X-ray.  The IR flaring has been described in \citet{chan2015b}.  The inclusion of non-thermal electrons in highly magnetized regions has produced significant X-ray variability which was previously unseen.}
\label{multi_variability}
\end{figure}

We find that X-ray variability is a generic result of constraining the non-thermal electrons to highly magnetized regions.  This is because the magnetic field is dynamic throughout the flow, generating magnetic flux tubes, which are in a constant state of being formed, sheared, becoming buoyant, and leaving the disk.  The dynamic nature of these flux tubes combined with strong lensing effects from the black hole generate both persistent variability as well as flaring events.  

X-ray flares in our simulations are always coincident with IR flares, but there are numerous IR flares without X-ray counterparts, as shown in Figure \ref{multi_variability}, which qualitatively matches observations. In this figure, we have zoomed in to the first 250 timesteps of the simulation in order to more clearly illustrate the relationship between the IR and X-ray lightcurves.   During this time span, we observe about 5 IR flares over the stochastically variable background and one significant X-ray flare.   From our simulation we find that there are about 5 IR flares per X-ray flare and a rate of one X-ray flare per 72,000 seconds, over the entire  simulation.  Over the course of 3 million seconds of observation with Chandra, 39 X-ray flares were observed, corresponding to one flare every $\sim$77,000 seconds.  Observations Sgr~A* show that, for every X-ray flare, there are about 4 NIR flares (e.g., \citealt{genzel2003,eckart2006}).  These numbers are in rough agreement with our results.

\begin{figure}[!h]
\includegraphics[width =0.5\textwidth]{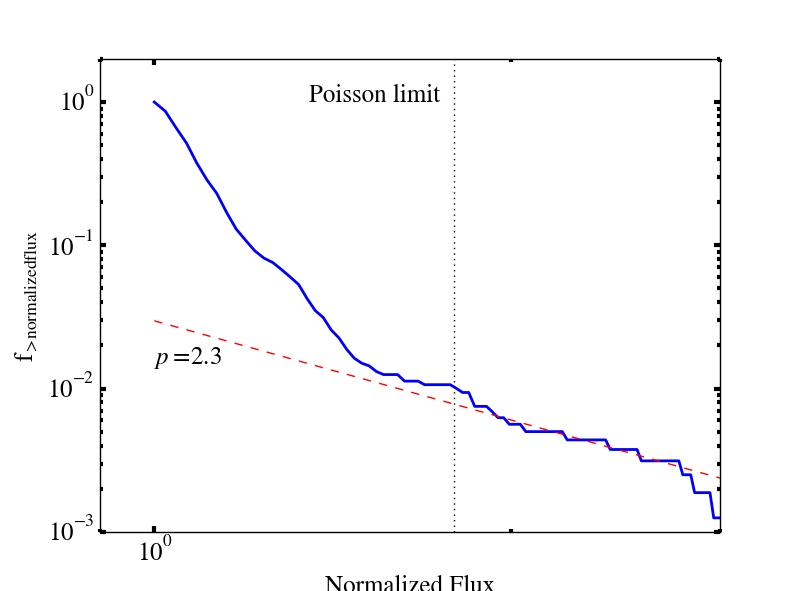}
\caption{X-ray flux distribution, accounting for a constant quiescent background.  At high fluxes, the flare distribution resemble a power-law with an index of $\sim$ 2.3.  Using the Poisson rate and binning reported in \citet{neilsen2015} for the Chandra observations used in that study, we estimate what would be the upper end of the Poisson-dominated regime, depicted with a vertical dotted line.   }
\label{flux_dist}
\end{figure}

In order to compare flare statistics from our simulations to observations more directly (e.g., through flux distributions reported in \citealt{neilsen2015}), we need to account for the fact that only $\sim$ 10\% of the X-ray emission from Sgr~A* comes from the inner accretion flow \citep{neilsen2013}.  In Figure \ref{flux_dist}, after adding a constant background equal to 90\% of the observed quiescent flux to our lightcurve from Figure \ref{therm_vs_nontherm}, we plot the flux distribution in our simulations.  We find that the flare distribution resembles a power-law with an index of around 2.3, while the lower-level variability does not have an obvious structure.  \citet{neilsen2015} reports only Poisson variability at low fluxes, and power-law behavior at high fluxes, with a power-law index of 1.92.  This is roughly consistent with our simulated flux distribution.  Our simulations, however, do not account for the Poisson photon counting noise, and also do not show as large of a range of variability.  The latter is likely due to the relatively short duration of our simulation that did not capture many rare, high flux events.

We estimate the level of the Poisson noise, below which we expect our simulated flux distribution to deviate significantly fom observations.  We take the reported photon counting rate ($Q=5.24$ cts/ks) and binning ($b=300$ s) from \citet{neilsen2015} and calculate the typical fractional Poisson error, $\epsilon =1/\sqrt{Qb}=.79$.  Normalizing our lowest level of emission to 1, we see that counting noise will dominate the observed variability from $1$ to $1+\epsilon$, setting the lower limit from which we expect our simulations to reproduce the flux distribution.

\begin{figure}[!h]
\includegraphics[width =0.45\textwidth]{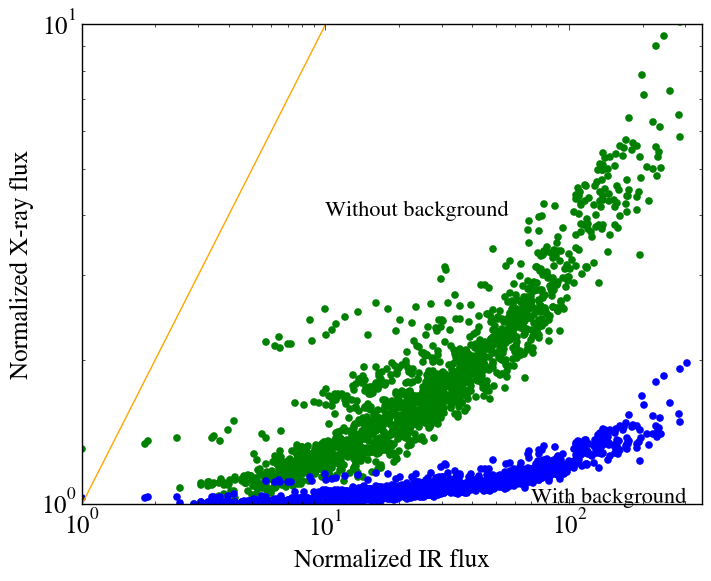}
\caption{IR vs. X-ray flux, accounting for a constant quiescent background  the of X-ray flux (blue), and purely for the inner accretion flow (green).  The orange line depicts a correlation with a slope of unity.  The addition of the quiescent background decreases the X-ray variability by a factor of $\sim$10.  We see a general trend of higher IR fluxes being associated with relatively high X-ray fluxes.}
\label{IR_X_scatter}
\end{figure}

We further explore the relationship between IR and X-ray fluxes in Figure \ref{IR_X_scatter}.  The largest IR flares correspond to the largest X-ray flares, but there is much more variability in the IR than the X-rays.  In our simulations, anytime a flux tube appears and crosses a caustic there will be an IR flare due to the synchrotron emission from the thermal electrons.  However, only the most highly magnetized flux tubes will have non-thermal electrons associated with them and will generate an X-ray flare. The effect of the $\beta$ threshold is to pick out a subset of all the magnetic flux tubes, ones with conditions suitable for reconnection to occur.  The particular threshold we use is motivated by \citet{liguo2015}, who showed a non-thermal component being generated for $\beta<0.2$.  As a result, IR variability is much more significant, since there is thermal synchrotron associated with all flux tubes, whereas particle acceleration and hence X-ray emission only occurs for a particular subset of the tubes.  Additionally, we see that the flux tubes responsible for the largest IR flares are the same structures responsible for the largest X-ray flares.  This is unsurprising given the strong scaling of synchrotron emissivity with magnetic field; the most highly magnetized flux tubes radiate copiously in the IR due to the high magnetic fields, and also act as sites of efficient reconnection, generating strong X-ray flares.

\section{conclusions}
In this paper, we explored the effects of incorporating non-thermal electrons into GRMHD simulations of radiatively inefficient accretion flows.  We found that X-ray variability is a generic result of constraining the non-thermal electrons to highly magnetized regions and that the timescales associated with electron cooling are comparable to the observed flare duration.
This analysis is model-dependent, since the synchrotron radiation from non-thermal electrons depends on the strength and topology of the magnetic field, which will impact the constraints we place on the quiescent energy budget of the non-thermal electrons.
Our analysis of cooling timescales are also likely to differ across models.  For model B with only inflowing velocities within 10 gravitational radii, we found that flares are unlikely to originate from the funnel since the inward radial velocities were too large to explain the X-ray flares lasting many thousands of seconds.  We will explore the role of non-thermal electrons in producing the variability for different magnetic field configurations and black hole spins in a future study.

In the context of model B, which has matched many observational constraints, from the quiescent broadband spectrum to the variability properties, we find that X-ray flares likely originate from magnetic flux tubes in the disk, where centrifugal support allows the non-thermal electrons to remain in the flow for many thousands of seconds while radiating away their energy via synchrotron emission.  The flare lengths are hence set by the synchrotron cooling timescales.  The properties of the X-ray variability from this model are consistent with observations: X-ray flares are always coincident with IR flares, there are many more IR flares without associated X-ray counterparts, and the timescales associated with the flares are comparable to the observed flare duration.

\acknowledgements

We thank Joey Neilsen, Daryl Haggard, Daniel Marrone, and Lia Medeiros for useful discussions. 
We gratefully acknowledge support for this work from Chandra Award No. TM6-17006X and from 
 NASA/NSF TCAN award NNX14AB48G.  
All ray tracing calculations were performed with the \texttt{El~Gato}
GPU cluster at the University of Arizona that is funded by NSF award
1228509.

\bibliographystyle{apj}
\bibliography{david_bib}

\end{document}